# A k-space method for nonlinear wave propagation

Yun Jing[a)] and Greg T. Clement

*Department of Radiology,*

*Harvard Medical School,*

*Brigham and Women's Hospital,*

*Boston 02115 USA*

(Dated: July 6, 2011)

**Abstract**

A k-space method for nonlinear wave propagation in absorptive media is presented. The Westervelt equation is first transferred into k-space via Fourier transformation, and is solved by a modified wave-vector time-domain scheme [Mast et al., IEEE Tran. Ultrason. Ferroelectr. Freq. Control 48, 341-354 (2001)]. The present approach is not limited to forward propagation or parabolic approximation. One- and two-dimensional problems are investigated to verify the method by comparing results to the finite element method. It is found that, in order to obtain accurate results in homogeneous media, the grid size can be as little as two points per wavelength, and for a moderately nonlinear problem, the Courant-Friedrichs-Lewy number can be as small as 0.4. As a result, the k-space method for nonlinear wave propagation is shown here to be computationally more efficient than the conventional finite element method or finite-difference time-domain method for the conditions studied here. However, although the present method is highly accurate for weakly inhomogeneous media, it is found to be less accurate for strongly inhomogeneous media. A possible remedy to this limitation is discussed.

## I. INTRODUCTION

A prevalent numeric approach to nonlinear acoustic problems involves solving the Kuznetsov-Zabolotskaya-Khokhlov (KZK) equation[1–7]. Developed as a modification of the Burgers equation to include absorption and diffraction[8], the KZK equation can also be derived as the parabolic approximation to the Westervelt equation[9]. Despite its utility, the equation is limited in validity to cases of quasi-planar wave propagation and is accurate for directional fields close to the axis of propagation and far from the source[10]. More general methods based on the Westervelt equation have been reported, but have primarily been limited to forward propagation[11], thus neglecting reverberation, or otherwise restricting nonlinear distortion to the normal direction[12–14].

Recently, several methods have been proposed to solve the Westervelt equation without such restrictions[15–17]. Likewise, the present work investigates a new wave-vector time-domain (k-space) based numerical algorithm[18–21] that can be applied to a wide range of nonlinear applications. The overall aim is to develop a computationally efficient method that is well-suited for heterogeneous media. The described approach is similar to the so-called pseudo-spectral method[22] in that both methods calculate the spatial differentiation in the k-space by Fourier transformation. However, the present method employs a straightforward time-stepping scheme that has been shown to be more accurate and less complex[19,20]. In addition, the approach is based on the second-order nonlinear wave equation, whereas the pseudo-spectral method[22] is based on a first-order nonlinear wave equation, which requires extra computation and storage of the particle displacement vector. As a result, the present method is less demanding of the storage and computation.

The paper proceeds as follows: Sec. II presents the derivation of the k-space method followed by the description of an absorbing layer devised to eliminate the problem of transformation-induced phase wrapping. Sec. III discusses simulation results for one-dimensional plane waves and several two-dimensional problems. Here, the validity of the

---

a)Electronic address: `jingy@bwh.harvard.edu`

k-space method in the context of nonlinear acoustics is verified. Criteria on choosing spatial and temporal step-size are also revealed. Sec. IV concludes the paper.

## II. THEORY

For a fluid medium with inhomogeneous acoustic properties, the nonlinear acoustic wave equation (Westervelt equation) is written as[7]

$$\rho \nabla \cdot (\frac{1}{\rho}\nabla p) - \frac{1}{c^2}\frac{\partial^2 p}{\partial t^2} + \frac{\delta}{c^4}\frac{\partial^3 p}{\partial t^3} + \frac{\beta}{\rho c^4}\frac{\partial^2 p^2}{\partial t^2} = 0, \quad (1)$$

where $p$ is the sound pressure, $c$ is the sound speed, $\delta$ is the sound diffusivity, $\beta$ is the nonlinearity coefficient, and $\rho$ is the ambient density. All material are assumed to be spatially varying functions. The equation inherently assumes a thermoviscous fluid, as the relaxation mechanism is not considered. However, the following algorithm can be readily modified to include power law absorption and dispersion[23].

By using the normalized wave field $f = \frac{p}{\sqrt{\rho}}$[19], the first-order derivative term is eliminated and the nonlinear wave equation becomes

$$\nabla^2 f - \frac{1}{c_0^2}\frac{\partial^2 f}{\partial t^2} = \sqrt{\rho}\nabla^2 \frac{1}{\sqrt{\rho}}f + \frac{1}{c_0^2}(\frac{c_0^2}{c^2}-1)\frac{\partial^2 f}{\partial t^2} - \frac{\delta}{c^4}\frac{\partial^3 f}{\partial t^3} - \frac{\beta}{\sqrt{\rho}c^4}\frac{\partial^2 f^2}{\partial t^2}, \quad (2)$$

where $c_0$ is the background speed of sound.

By defining an auxiliary field $w = f + v$ where $v = (\frac{c_0^2}{c^2}-1)f$, Eq. (2) can be reduced to

$$-\frac{1}{c_0^2}\frac{\partial^2 w}{\partial t^2} = \nabla^2 v - \nabla^2 w + (q - h - d)/c_0^2, \quad (3)$$

where

$$q = c_0^2\sqrt{\rho}\nabla^2 \frac{1}{\sqrt{\rho}}f,$$
$$h = c_0^2\frac{\beta}{\sqrt{\rho}c^4}\frac{\partial^2 f^2}{\partial t^2},$$
$$d = c_0^2\frac{\delta}{c^4}\frac{\partial^3 f}{\partial t^3}. \quad (4)$$

Fourier transformations of Eq. (3) in the spatial domain yields the k-space equation

$$\frac{\partial^2 W}{\partial t^2} = c_0^2 k^2(V - W) - (Q - H - D), \quad (5)$$

where $k = \sqrt{k_x^2 + k_y^2 + k_z^2}$, $W, V, Q, H$, and $D$ are the spatial Fourier transform of $w, v, q, h$ and $d$, respectively, which can be calculated using a fast Fourier transform.

Equation (5) can be solved in a nonstandard finite difference approach[19,20],

$$W(t + \triangle t) - 2W(t) + W(t - \triangle t) = 4 sin^2(\frac{c_0 k \triangle t}{2}) \left[V - W - \frac{1}{c_0^2 k^2}(Q - H - D)\right], \quad (6)$$

where $\triangle t$ is the temporal step-size. For a homogeneous medium or a weakly inhomogeneous medium, this harmonic oscillator equation guarantees small dispersion error with large temporal step as opposed to the conventional leap-frog scheme,

$$W(t + \triangle t) - 2W(t) + W(t - \triangle t) = (c_0 k \triangle t)^2 \left[V - W - \frac{1}{c_0^2 k^2}(Q - H - D)\right], \quad (7)$$

or even higher order time integration, such as the fourth-order Adams-Bashforth time integration[22].

Comparing Eq. (6) with Eq. (9) in Ref.19, two new terms, $H$ and $D$ have been added, which are the contribution from the nonlinearity and diffusivity. This approach can be then viewed as a straightforward extension to a previous k-space method described by Mast, *et. al*[19] with nonlinearity and absorption included. In addition, since the coefficients of third and fourth terms in Eq.(1) are small compared with the coefficient of second term, the stability condition should be similar to the linear k-space method,

$$sin(\frac{\text{CFL} c_0 \pi}{2 c_{max}}) \leq \frac{c_0}{c_{max}}, \quad (8)$$

where $CFL$ is the Courant-Friedrichs-Lewy number $c_{max} \triangle t / \triangle x$ (Note that the definition is somewhat different from that used in Ref.19). Clearly, the algorithm is unconditionally stable for media with $c(\boldsymbol{r}) < c_0$ everywhere.

To calculate $H$ (or $h$), the following backward difference approximation was employed,

$$\frac{\partial^2 f^2}{\partial t^2} = \frac{45 f^2(t) - 154 f^2(t - \triangle t) + 214 f^2(t - 2\triangle t) - 156 f^2(t - 3\triangle t) + 61 f^2(t - 4\triangle t) - 10 f^2(t - 5\triangle t)}{12 \triangle t^2}.$$
$$(9)$$

This backward difference approximation has a fourth order accuracy. It has the disadvantage of requiring the storage of six time steps. However, if this is problematic, the second order backward difference approximation can be used at the expense of a significantly reduced time step.

$$\frac{\partial^2 f^2}{\partial t^2} = \frac{2f^2(t) - 5f^2(t-\triangle t) + 4f^2(t-2\triangle t) - f^2(t-3\triangle t)}{\triangle t^2}. \quad (10)$$

Conversely, if the memory is not limited, a higher order, e.g., a sixth order approximation might be a good alternative. In this study, however, only the fourth order approximation was used. The pressure field for the initial six steps can be either roughly calculated using the linear k-space method assuming that the nonlinearity does not significantly build up in these six steps, or more precisely calculated using nonlinear projection methods. In this study, the former approach was used.

Similarly, to calculate $D$ (or $d$), the following third-order backward difference approximation was used,

$$\frac{\partial^3 f}{\partial t^3} = \frac{17f(t) - 71f(t-\triangle t) + 118f(t-2\triangle t) - 98f(t-3\triangle t) + 41f(t-4\triangle t) - 7f(t-5\triangle t)}{4\triangle t^3}. \quad (11)$$

Since fast Fourier transform implies periodicity, the k-space method inherently has a wrapping artifact issue: the wave enters one boundary and exits from the other side. The easiest way to overcome this is to enlarge the computational domain, however, this inevitably increases the calculation time. The well-known perfectly matched layer technology (PML)[20] has been shown to be a good solution to this problem but is limited to the first order acoustic equation. In this study, an absorbing layer[24,25] was used to minimize the spurious reflections from the boundary. The equation for the absorbing layer can be written in $f$ as

$$\nabla^2 f - \frac{1}{c_0^2}\frac{\partial^2 f}{\partial t^2} = 2\gamma\frac{\partial f}{\partial t} + \gamma^2 f. \quad (12)$$

It is noted that the nonlinearity and absorption terms should still be considered in the absorption layer to prevent sudden medium change. They are only ignored here to emphasize the newly added term in the absorption layer, i.e., $\gamma$, which is an absorption term (frequency independent) and its derivative should be kept as small as possible. In this study, we have used[25]

$$\gamma = U_0/cosh^2(\alpha n), \tag{13}$$

where $U_0$ is a constant (2.0 in the present study), $\alpha$ is a decay factor (0.1 in the present study), and $n$ denotes the distance in number of grid points from the boundary. Although this equation was proposed for the linear acoustic equation, it has been found in this study to be also suitable for the nonlinear acoustic equation. Numerical simulations have shown that, for a normal incident wave, the reflection from the absorption layer can be reduced by more than $50dB$.

After applying the Fourier transform, Eq. (12) can be written in terms of $W$ as

$$\frac{\partial^2 W}{\partial t^2} = c_0^2 k^2(-W) - M, \tag{14}$$

where $M$ is the Fourier transform of $m$, and

$$m = c_0^2(2\gamma\frac{\partial f}{\partial t} + \gamma^2 f). \tag{15}$$

The first order time derivative can be calculated by the second order approximation

$$\frac{\partial f}{\partial t} = \frac{3f(t) - 4f(t - \triangle t) + f(t - 2\triangle t)}{2\triangle t}. \tag{16}$$

A higher order approximation can also be employed, but in this study, Eq.16 was found to be sufficient. Finally, Eq. (14) can be solved in a similar way as in Eq. (6).

To close this section, it is reiterated that the present nonlinear k-space algorithm is very similar to the linear k-space algorithm and can now be summarized as follows:

Step 1: set any initial conditions for $f$ (spatially smooth) for the first six time steps, this leads to $w$, and spatially Fourier transform to obtain initial conditions for $W$.

Step 2: compute $v$ and transform to obtain $V$, compute $q$ and transform to $Q$.

Step 3: compute the third, second, and first time derivative of $f$ to obtain $h, d$ and $m$, then transform to $H, D$, and $M$.

Step 4: evaluate $W(t + \triangle t)$ by Eq. (6) and inverse transform to obtain $w(t + \triangle t)$, which leads to $f(t + \triangle t)$

Step 5: set $t \to t + \triangle t$ and go to step 2

## III. SIMULATIONS

### A. A one-dimensional homogeneous medium problem

To verify the present k-space algorithm, as well as to determine the requirement for the spatial and temporal steps to achieve accurate results, we initially tested one-dimensional propagation in a homogeneous medium. The incident wave $p_i$ was defined as a plane wave with Gaussian temporal shape[19]:

$$p_i = p_0 sin(\omega_0 \tau)e^{-\tau^2/(2\sigma^2)}, \qquad (17)$$

where $p_0$ is the pressure amplitude, $\omega_0$ is the center angular frequency, $\tau$ is the retarded time $\tau = t - (x - x_0)/c_0$ and $x_0$ is the center of the pulse. In this section, $p_0$ was chosen to be $1MPa$, $\omega_0$ was $0.2MHz \times 2\pi$, and $\sigma$ was $10^{-10}$. This gave a nominal maximum frequency of $0.3MHz$, corresponding to the spectral point $70dB$ down from the center frequency. The speed of sound was $1500m/s$, the density was $1000kg/m^3$, and the nonlinearity was 3.5. No attenuation was considered. With Eq. (17), the sound field is known for the first six temporal steps, with the assumption that the nonlinearity is negligible for these six steps. This accomplishes Step 1 in the algorithm described in the previous section.

A benchmark solution was also obtained using the finite element method[26], to simulate the Westervelt equation. Quadratic Lagrange-type elements were used. The interpolating polynomial used in the time-stepping method was the second-order backward differentiation formula. To guarantee the accuracy of the benchmark solution, up to 80 elements per wavelength (at the center frequency) were used, and the CFL number was kept 0.1. Convergence study was carried out and this configuration has been found to be sufficient for moderately nonlinear problems.

For the k-space algorithm, the spatial step was chosen to be $1/2, 1/4, 1/6$, and $1/8$ of the wavelength at the nominal maximum frequency. The CFL number was also varied from 0.1 to 0.4, with an increment of 0.1. Figure 1a shows the time-domain signal of the pulse after a distance of $22.5cm$, which corresponds to 0.3 of the theoretical plane wave shock formation distance for a sinusoidal wave at $0.2MHz$. In this case, the CFL number was fixed at 0.1, but the spatial step varied. It can be observed that the k-space results are indistinguishable from the finite element method results, except for the lowest spatial resolution, 1/2 wavelength. Figure 1b further compares the results in the frequency domain, where the difference becomes more pronounced and illustrative. The k-space method is valid approximately up to $0.3MHz, 0.6MHz, 0.9MHz$ and $1.2MHz$ for the step size of $1/2, 1/4, 1/6$ and $1/8$ wavelength, respectively, due to the Nyquist rate theory. Therefore, in a weakly nonlinear problem, where the fundamental and second harmonics are of concern, the k-space method only needs to employ 4 points per wavelength at the maximum frequency to capture accurate wave propagation for a homogeneous medium.

The temporal criteria is also tested by fixing the step size to 1/8 of the wavelength and varying the CFL number from 0.1 to 0.4. Figure 1c shows the time-domain signal at the same distance. Figure 1d shows the frequency spectrum. As expected, the smaller the CFL number is, the more accurate the result is. For CFL number $0.1, 0.2$ and $0.3$, the difference is almost negligible over the entire frequency domain. For CFL number 0.4, the difference becomes only noticeable when the frequency goes beyond $0.8MHz$. Therefore, for such a moderately nonlinear problem, a CFL number as high as 0.4 can be used for reasonably accurate results (errors of the fundamental and first two harmonics being less than $0.5dB$). It is expected that for a weakly nonlinear problem, the required CFL number can be even larger than 0.4, because the nonlinearity can then be viewed as a small perturbation imposed upon the linear wave equation. For a linear problem, the CFL number in the k-space algorithm can be arbitrarily large in a homogeneous medium as long as the Nyquist rate is satisfied.

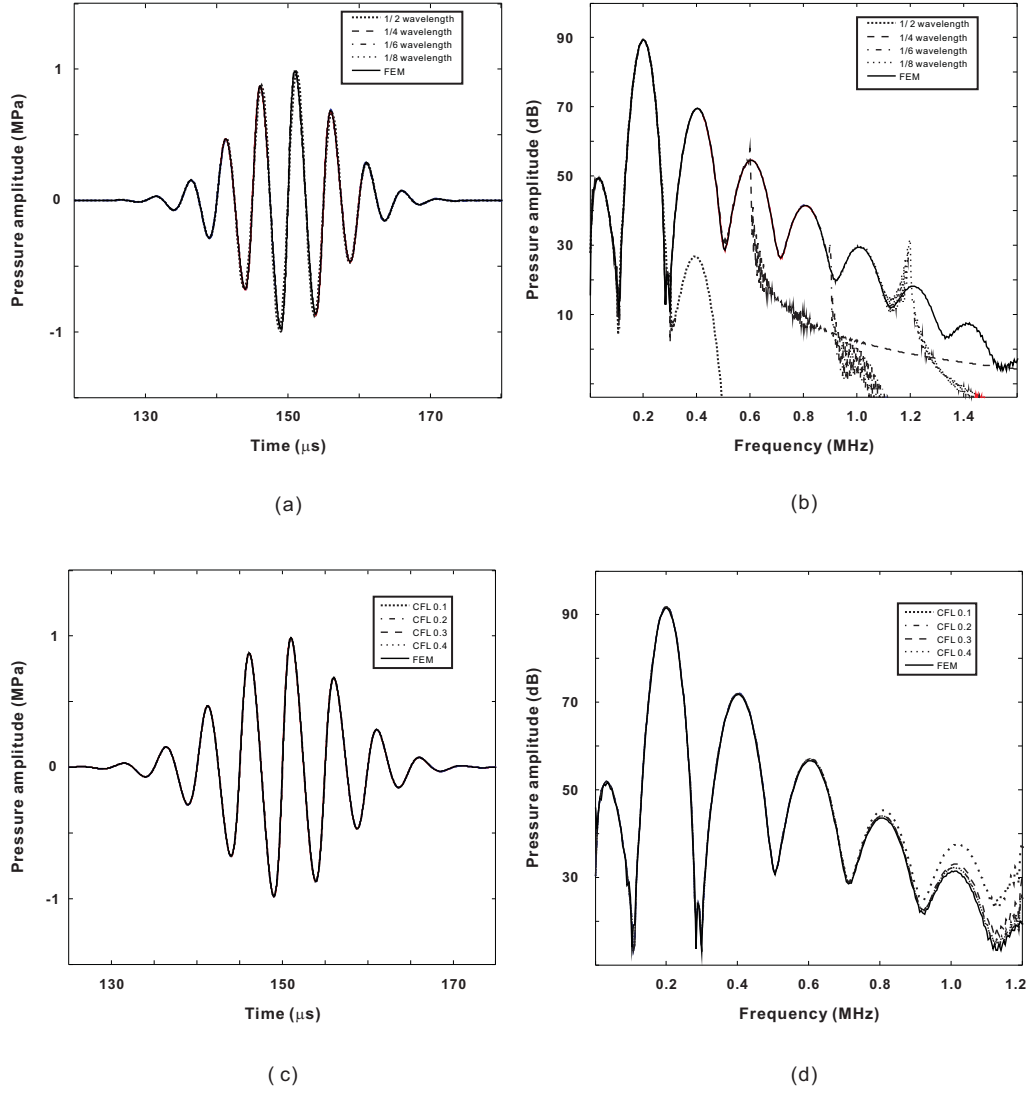

FIG. 1. Comparison between the finite element method and nonlinear k-space method for a one-dimensional homogeneous medium at a distance of approximately 0.3 of the shock formation distance. (a) Time-domain results with spatial step varied. (b) Frequency-domain results with spatial step varied. (c) Time-domain results with CFL number varied. (d) Frequency-domain results with CFL number varied.

## B. A one-dimensional inhomogeneous medium problem

The purpose of this example is to test the k-space method for a one-dimensional inhomogeneous medium. The pulse described in the preceding section was used, with the pulse centered at $x = -10 cm$: $c = 2250 m/s$, $\rho = 1200 kg/m^3$, and $\beta = 2$. For the k-space algorithm, spatial step was 1/6 of the wavelength at the maximum frequency, and the CFL number was 0.3.

Figure 2a shows the time domain signal at $x = 12.5 cm$, showing the wave at the same distance as that shown in the preceding section. No significant difference can be observed between the result of finite element method and the k-space method. Figure 2b shows the frequency spectrum. Excellent agreements are obtained at the fundamental and second harmonic frequencies. For the third harmonic the difference is about $0.5 dB$, for the fourth is about $2.0 dB$. A previous study has shown that for linear inhomogeneous media, 2 cells per wavelength are insufficient for accurate results[18], as any jump of the medium property represents high spatial frequency and requires many more cells per wavelength. To consider harmonics in a nonlinear problem, $2N$ cells per wavelength are required ($N$ is the number of harmonics considered). When $N$ is high, it automatically improves the accuracy of the k-space method for modeling wave propagation in inhomogeneous media.

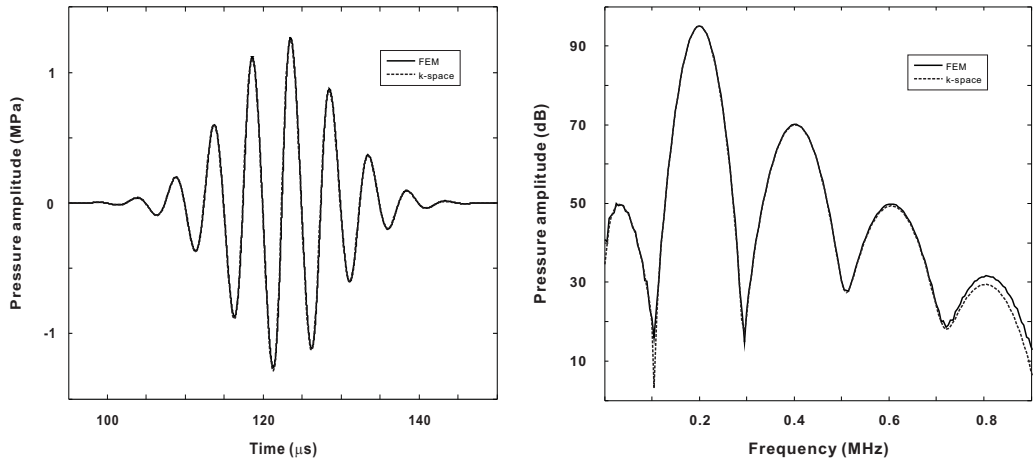

FIG. 2. Comparison between the finite element method and nonlinear k-space method for a one-dimensional inhomogeneous medium. (a) Time-domain results. (b) Frequency-domain results.

## C. A two-dimensional homogeneous medium problem

This sections discusses nonlinear wave propagation in a two-dimensional homogeneous medium. The medium acoustic property remains the same as in the preceding one-dimensional homogenous problem, except that the nonlinearity coefficient was 4.0. The initial sound field again employed Eq. (17), where the center frequency was $1MHz$, $x_0 = -0.012, \sigma = 3e^{-12}$, resulting in a nominal maximum frequency of $1.5MHz$. The pressure amplitude $p_0$ was $2MPa$. In addition, this initial condition only applied to the domain $|y| \leq 0.01$. For the FEM simulation, the calculation domain was $4.5cm \times 4.5cm$, where approximately 12 elements per minimum wavelength were used. The CFL number was 0.25. A first order non-reflecting boundary condition[27] was used to eliminate spurious reflections. For the k-space simulation, the calculation domain was $5cm \times 5cm$, where 6 elements per minimum wavelength were used. The CFL number was 0.3.

Figure 3 illustrates three sets of snapshots of the sound field at time $6.23\mu s$, $9.80\mu s$, and $13.38\mu s$. The upper three figures show the FEM results, and the lower three figures show the k-space results. The area shown in each figure is $2.5cm \times 2.5cm$. Visually there is only a slight difference between the FEM and k-space results near the bottom and the two side boundaries, due to the flaw of the first order non-reflecting boundary condition used in the FEM simulation. To quantify the difference between these two, the least-square error was used as defined by[19]

$$error = \frac{\|p_{k-space}(\boldsymbol{r}) - p_{FEM}(\boldsymbol{r})\|}{\|p_{FEM}(\boldsymbol{r})\|}, \qquad (18)$$

where $\|p(\boldsymbol{r})\|$ is the least-square norm.

The least square errors for the three different times are reasonably small and are $0.0333, 0.0301$, and $0.0315$, respectively.

Figure 4(a) shows both FEM and k-space results at the location $(8.25, 0)mm$ in the time-domain, while Fig. 4(b) shows the results in the frequency-domain. It can be seen that the k-space result is in good agreement with the FEM result.

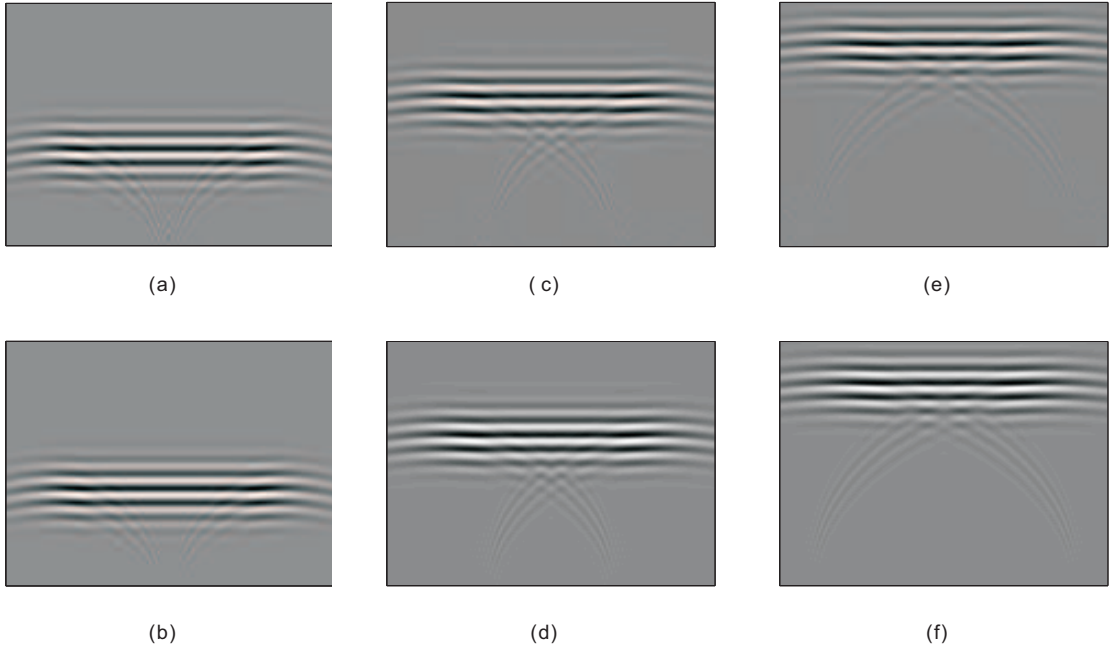

FIG. 3. Sound field computed in a two-dimensional homogeneous medium. Panels (a), (c) and (e) show the pressure at $6.23\mu s$, $9.80\mu s$, and $13.38\mu s$ obtained from the FEM. Panels (b), (d) and (f) show the pressure at $6.23\mu s$, $9.80\mu s$, and $13.38\mu s$ obtained from the k-space.

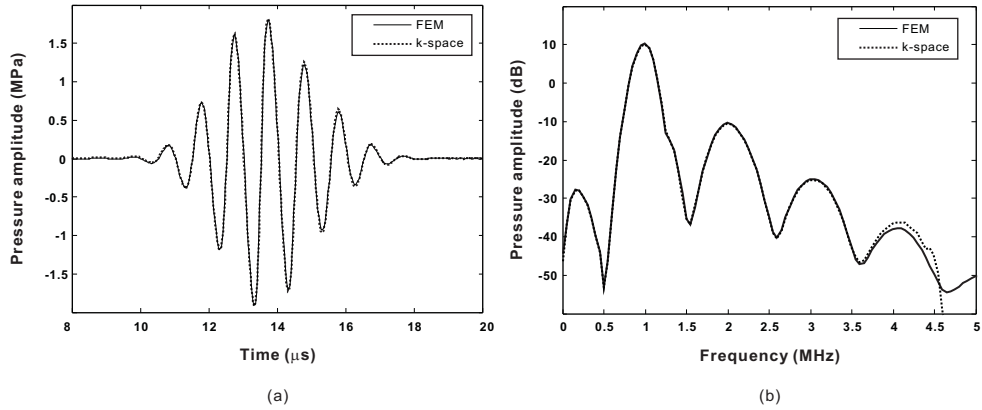

FIG. 4. Sound field computed in a two-dimensional homogeneous medium. (a)Time-domain results from k-space and FEM at the location $(8.25, 0)mm$. (b) Frequency-domain results at the same location.

**D. Two two-dimensional inhomogeneous medium problems**

In this section, two two-dimensional inhomogeneous medium problems are discussed. In the first problem, a cylindrical object with radius $4.0mm$ was added into the center of the medium as a scatterer. The background medium has the same acoustic property in the one-dimensional homogenous medium problem, i.e., the speed of sound was $1500m/s$, the density was $1000kg/m^3$, and the nonlinearity was 3.5. The acoustic property in the cylindrical object was: the speed of sound was $1575m/s$, the density was $1050kg/m^3$, and the nonlinearity was 4. Clearly, this is a weak contrast problem, which can occur for example in soft tissue[11]. The initial sound field, as well as the algorithm for both methods was the same as in the preceding section. Figure 5 presents three sets of snapshots of the sound field at time $6.23\mu s$, $9.80\mu s$, and $13.38\mu s$. The upper three figures show the FEM results, and the lower three show the k-space results. The area shown in each figure is $2.5cm \times 2.5cm$. The least square errors are $0.0339, 0.0320$, and $0.0334$, respectively. For this weakly inhomogeneous problem, the k-space method is shown to have fairly good performance, as was the case for homogeneous problems. This has been found also in linear acoustic problems[18–20].

Figure 6(a) illustrates both FEM and k-space results at the location $(8.25, 0)mm$ in the time-domain, while Fig. 6(b) shows the results in the frequency-domain. Again, the k-space result agrees well with the FEM result.

In the second example, the acoustic property in the cylindrical object was changed to: the speed of sound was $3000m/s$, the density was $2000kg/m^3$, and the nonlinearity was kept 4. This example allowed to verify the present method for strong contrast problems, which can occur in for example phononic band gaps[28]. In the same fashion, Fig. 7 presents three sets of snapshots of the sound field at time $6.23\mu s$, $9.80\mu s$, and $13.38\mu s$. The upper three figures show the FEM results, and the lower three figures show the k-space results. The area shown in each figure is $2.5cm \times 2.5cm$. The least square errors appear to grow and are $0.0795, 0.102$, and $0.102$, respectively. As also discovered in linear acoustic problems[18–20], the k-space method becomes less accurate for strongly inhomogeneous problems. Figure

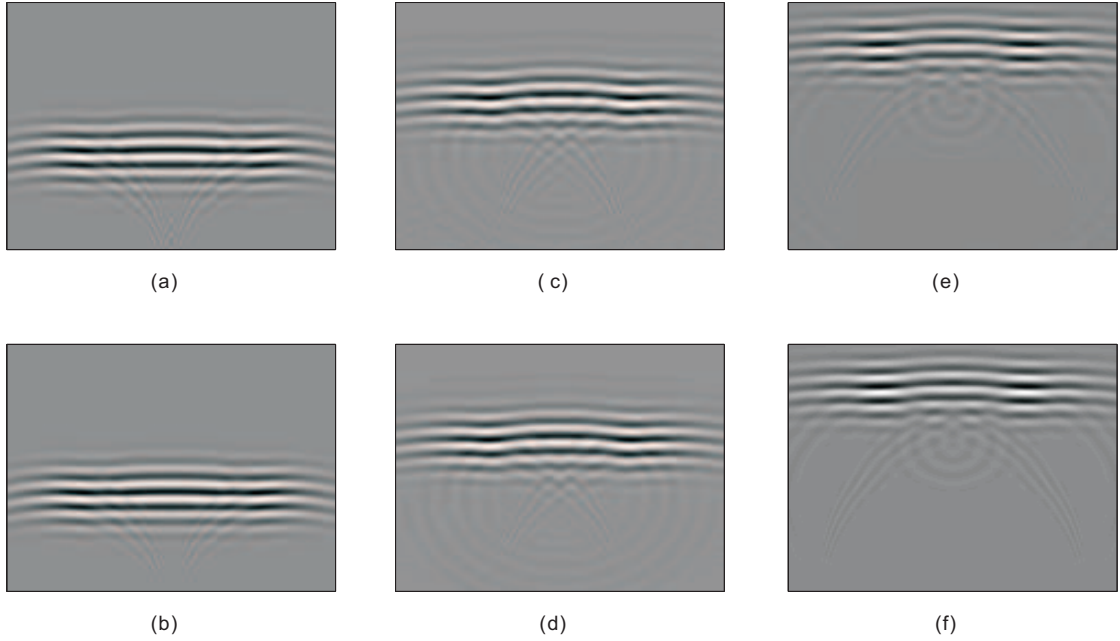

FIG. 5. Sound field computed in a two-dimensional weakly inhomogeneous medium. Panels (a), (c) and (e) show the pressure at $6.23\mu s$, $9.80\mu s$, and $13.38\mu s$ obtained from the FEM. Panels (b), (d) and (f) show the pressure at $6.23\mu s$, $9.80\mu s$, and $13.38\mu s$ obtained from the k-space.

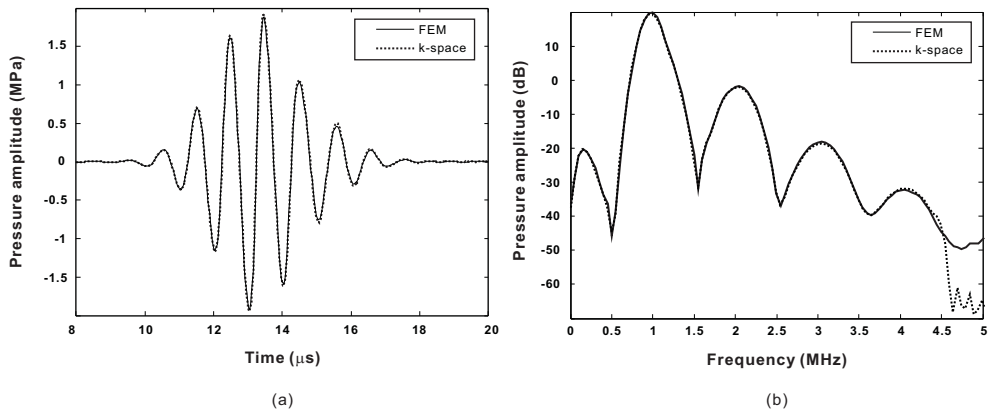

FIG. 6. Sound field computed in a two-dimensional weakly inhomogeneous medium. (a)Time-domain results from k-space and FEM at the location $(8.25, 0)mm$. (b) Frequency-domain results at the same location.

8(a) shows both FEM and k-space results at the location $(8.25, 0)mm$ in the time-domain,

while Fig. 8(b) shows the results in the frequency-domain.

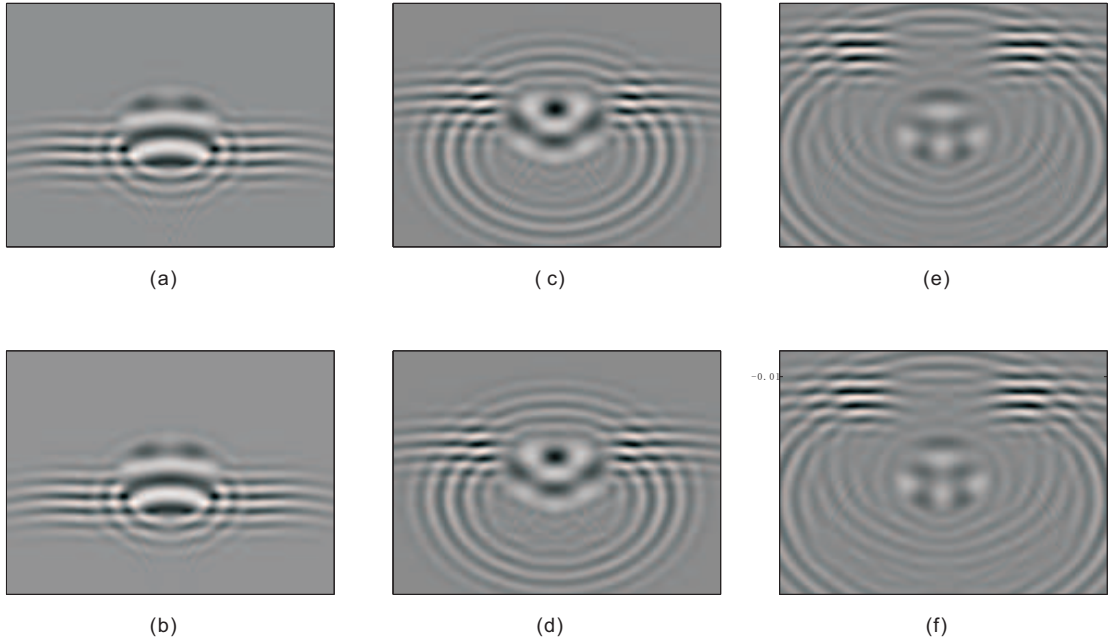

FIG. 7. Sound field computed in a two-dimensional strongly inhomogeneous medium. Panels (a), (c) and (e) show the pressure at $6.23\mu s$, $9.80\mu s$, and $13.38\mu s$ obtained from the FEM. Panels (b), (d) and (f) show the pressure at $6.23\mu s$, $9.80\mu s$, and $13.38\mu s$ obtained from the k-space.

### E. Attenuation effect

This section discusses effects of attenuation, i.e., the influence of the term $\delta$ on the nonlinear acoustic field. The problem under test follows the description in Sec. III. C, the homogeneous two-dimensional case, but with attenuation added ($\delta = 1 \times 10^{-3} m^2 s^{-1}$). This attenuation corresponds to approximately $0.51 dB/cm$ at $1MHz$. Numerical algorithm was kept the same as in Sec. III. C. Figure 9 shows the result at the point $(8.25, 0)mm$ in the frequency domain. Time-domain results are not presented because there is no visibly apparent discrepancy. The k-space method agrees with the FEM results well up to the 3rd harmonics. The k-space method results without attenuation are also provided for reference.

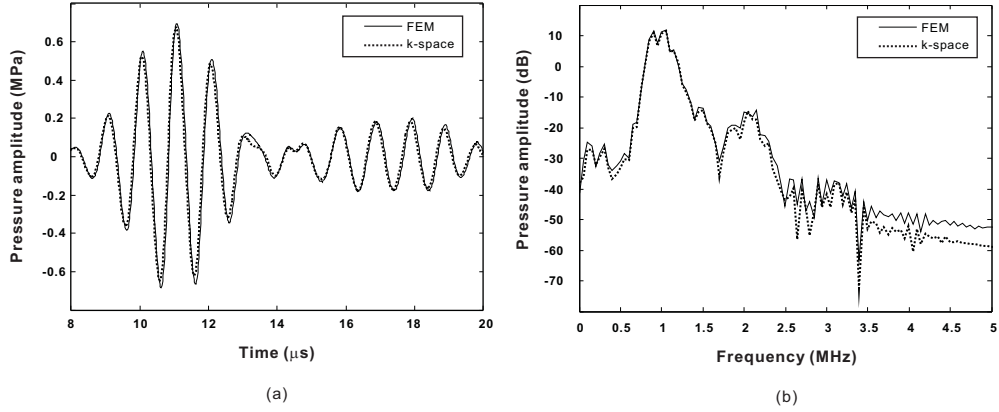

(a)　　　　　　　　　　　　　　(b)

FIG. 8. Sound field computed in a two-dimensional strongly inhomogeneous medium. (a)Time-domain results from k-space and FEM at the location $(8.25, 0)mm$. (b) Frequency-domain results at the same location.

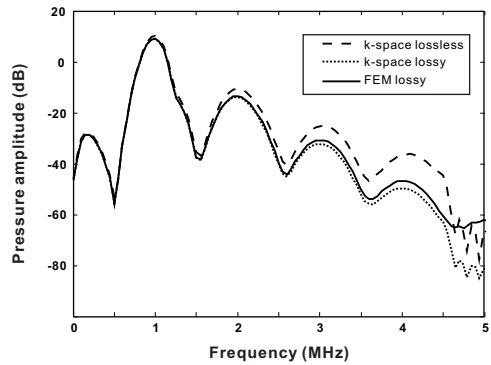

FIG. 9. Sound field computed in a two-dimensional homogeneous medium with attenuation considered. Frequency-domain results at the location $(8.25, 0)mm$ are shown.

## IV. CONCLUSIONS AND DISCUSSIONS

This paper reports on a newly developed algorithm for nonlinear wave propagation based on the k-t space scheme[19]. The validity of the present method is tested by comparing it with a finite element method, where good agreements have been found. The present approaach can be viewed as an extension to a previously reported linear k-space method[19], which has been expanded to include the effects of nonlinearity and attenuation. The method takes inhomogeneity into account and, unlike standard methods, does not employ the parabolic approximation or assume one-way propagation. Thus it can be applied to a wide range of nonlinear acoustic problems. Furthermore, many commonly used approaches assume nonlinear distortions occur normal to the source plane, which is not a valid assumption for highly focusing transducers. This method solves the full Westervelt equation without neglecting the nonlinear distortions along other directions, and thus might be a useful tool for studying the nonlinear sound field for highly focusing transducers. The present method is highly efficient, as it is a spectral approach, and thus only requires two nodes per minimum wavelength, as set by the Nysquist rate. In addition, since the method uses an extremely accurate time-stepping algorithm[19], the temporal step (or CFL number) can be fairly large compared with other approaches. As shown in the simulation results, accurate results can be obtained using the present method for homogeneous or weakly inhomogeneous media. Less accurate results were obtained in a strongly inhomogeneous media. It is expected that this is caused in part by aliasing, which occurs from Fourier transformation of discontinuities in the medium[20]. In addition, the second-order wave equation incorporates the density within a second-order derivative term, which can be hard to represent numerically[20]. A possible way to fix this would be to employ the coupled first-order nonlinear wave equations[20,22], which have already shown significant improvement for linear wave propagation in inhomogeneous media.

## V. ACKNOWLEDGEMENTS

The authors would like to thank Dr. Ben Cox for helpful discussions. This work was supported, in part, by NIH grant R01EB003268.

**List of Figures**